ORCID #: 0000-0003-2234-8756

# Simulating the Ridesharing Economy: The Individual Agent Metro-Washington Area Ridesharing Model (IAMWARM)

**Joseph A. E. Shaheen**

jshaheen@gmu.edu
Computational and Data Science Department, George Mason University
4400 University Drive, MS 6A2
Fairfax, VA 22030-4444

## Abstract

The ridesharing economy is experiencing rapid growth and innovation. Companies such as Uber and Lyft are continuing to grow at a considerable pace while providing their platform as an organizing medium for ridesharing services, increasing consumer utility as well as employing thousands in part-time positions. However, many challenges remain in the modeling of ridesharing services, many of which are not currently under wide consideration. In this paper, an agent-based model is developed to simulate a ridesharing service in the Washington D.C. metropolitan region. The model is used to examine levels of utility gained for both riders (customers) and drivers (service providers) of a generic ridesharing service. A description of the Individual Agent Metro-Washington Area Ridesharing Model (IAMWARM) is provided, as well as a description of a typical simulation run. We investigate the financial gains of drivers for a 24-hour period under two scenarios and two spatial movement behaviors. The two spatial behaviors were random movement and Voronoi movement, which we describe. Both movement behaviors were tested under a stationary run conditions scenario and a variable run conditions scenario. We find that Voronoi movement increased drivers' utility gained but that emergence of this system property was only viable under variable scenario conditions. This result provides two important insights: The first is that driver movement decisions prior to passenger pickup can impact financial gain for the service and drivers, and consequently, rate of successful pickup for riders. The second is that this phenomenon is only evident under experimentation conditions where variability in passenger and driver arrival rates are administered.

# 1. Introduction

Ridesharing services such as Uber and Lyft have been experiencing explosive growth driven by rider demand and a number of other factors in industry [1]. One challenge for both ride-sharing service providers as well as for their driver contractors is how to maximize driver acceptance of new customers while ensuring drivers gain maximum utility from rides given. In other words, how to maximize both financial gains for the service and its drivers while ensuring maximum service-level quality for its customers.

Perhaps as a reflection of the growth of the ridesharing industry and of the aforementioned challenges, empirical research in this area is also experiencing a surge, exemplified by a growing number of journal publications [1-18] that explore the multidimensional challenges and opportunities produced by the widespread adoption of ridesharing services.

This paper aims to investigate spatial behavioral conditions under which drivers can gain increased financial returns (utility) on their invested time, while simultaneously ensuring that a maximum number of potential passengers reach their destination. The model produced utilizes an agent-based modeling (ABM) framework that has the potential to be extended, expanded, and tested under many variable conditions. And therefore, while it would be immediately salient that much can and should be tested with the model, we reserve future extensions and testing for future papers.

It is worthwhile to note that the agent simulation perspective is highly suitable for testing spatial behaviors; when utilizing agent simulations it is considered trivial to create many autonomous, heterogeneous agents following one or more behavioral rule-sets and to simulate behaviors for various initial conditions and parameters without the constraints of rigid assumptions. Additionally, as will be shown by the results of our experimentation, other modeling techniques may not fully capture the true temporal dynamics of a ridesharing service because of heterogeneity in agent decision-making, the spatial significance to end results, and the variable scenario conditions under which emergent properties could arise. In a subsequent section, we will show the relevance of the later.

The model described in this paper focuses on simulating drivers and riders in the Washington, D.C. metro region and attempts to simulate the movement of drivers under two spatial movement conditions.

Ultimately, the aim of this model is to gain insight into whether drivers, riders and ridesharing services benefit more or less from optimized decision-making during the drive-pickup-drop-off lifecycle familiar to ridesharing customers, while the aim of this paper is to highlight testing of some specific conditions.[1] In later iterations of the model, an increased variety of behaviors will be investigated.

### *1.1* Background

There have been a number of studies in the last few years taking aim at understanding ride-sharing services and carpooling schemes – each of which takes a different investigative position on the challenges faced by resource pooling services as a whole [1, 22], while some consider some facet of modeling behaviors [3] using agent-based approaches. The majority of papers reviewed were of modeling carpooling decisions as an optimization problem [7, 11, 12] and finally, some approaches intended to make early-stage predictions about carpooling and ridesharing trends [6, 9] were pre-existent in the literature.

What becomes very clear during a topical literature review over the last few years is that no real attempt had been made to provide for a comprehensive ridesharing agent-based simulation that captures prevalent dynamics; though much of the research attempts to understand the effects of ridesharing in general.

For example, Cho et. al. [3] provided a full description of a hypothetical agent-based model for a carpooling application without offering an actual build of the model hypothesized. The authors focused on the systemic theoretical structure of the proposed model, the mathematics and optimization techniques that *would* be used and the general form of social network types that *could* be used between the driver agents of said model. The same group [4] later proposed another agent-based model – this time only based on social network interactions without implementation.

Significant advances in the area of heuristics and algorithm development that propose better route optimization techniques have also been made over the last few years and this is an area where high-value and productive work has been put forward. For example, Pelzer et. al. [12] developed a method which aims to best utilize "ridesharing potential while keeping detours below a specific limit" using a spatial partitioning method.

IAMWARM aims to build a foundational baseline to test a small number of interesting spatial problems for which answers have not been provided as of yet and to use the model created as the basis for future improvements, extensions, expansions, and experiments. We begin that endeavor by discussing our primary and most central question: Can ridesharing utility for both riders, drivers, and service be increased through varying the information-shared among agents, ultimately affecting spatial movement behaviors? And if so, under what conditions could one note a difference in system-level properties?

---

[1] The author of this paper registered with one ridesharing service in order to gain insight into the natural behaviors of drivers and riders of the service. A total of 30 trips were carried out.

### *1.2* Information Asymmetry vs Information Symmetry

Our path is to find the simplest method of testing ridesharing utility schemes spatially, so we begin by discussing the problem of increased information sharing briefly implied in the previous section.

For our specific context, we define information-asymmetry as a lack of information regarding the location of other drivers by other drivers. That is—drivers, in an information-asymmetric service, would not be given the locations of other drivers, or riders[2], with exception of a single potential rider within their vision's radius who has just requested a pick-up, and thus without that information and without clear route planning driven by spatial demographics, drivers would simply move about randomly hoping to 'luck out' and be near a potential customer when they request a pickup. This is currently the method by which all ridesharing services manage their respective platforms. Drivers of those services are not given location information of other drivers, and must move about based on randomness, their own past experiences and information gleaned from their social networks; and so ultimately, must make their spatial movement and positioning decisions based on either luck or experience gained from learning. We will omit learning behavior from this iteration of the model.

Symmetry represents a condition such that driver agents have all the available information about other driver agents and rider agents. For this model however we bound true information symmetry to a localized version that limits driver agents' knowledge to the nearest driver agent and only to the nearest rider agent. The comparison between information asymmetry and symmetry which will be established by the comparison between random movement and Voronoi movement will be applied such that agents have no vision for information asymmetry scenario runs (no knowledge of the position of any other driver) and have only local vision in the information symmetry variation of the model (knowledge of the nearest driver agent's position). This modification is a direct result of the platform chosen for the development of the simulation and its ability to perform, and due to a lack of a clear theoretical or even observed cognitive standard to base spatial behaviors upon in this case.

This model will investigate how drivers would benefit from having local information about peer drivers available in real time using a hypothesized spatial behavior and whether that additional information would maximize driver utility. And, though there are a number of differing spatial behaviors that can be considered, we will only test one behavior which we theorize would result from access to that information. We will extend this model in future iterations with more behaviors.

[2] We will later explain our terminology in detail, but for now we define a driver as an agent who is picking up a rider from one location on our model's spatial grid to another. Once a rider is "picked up" we will refer to him as a passenger. In our model a passenger is no longer an agent but is a data point in the driver agent's attributes list.

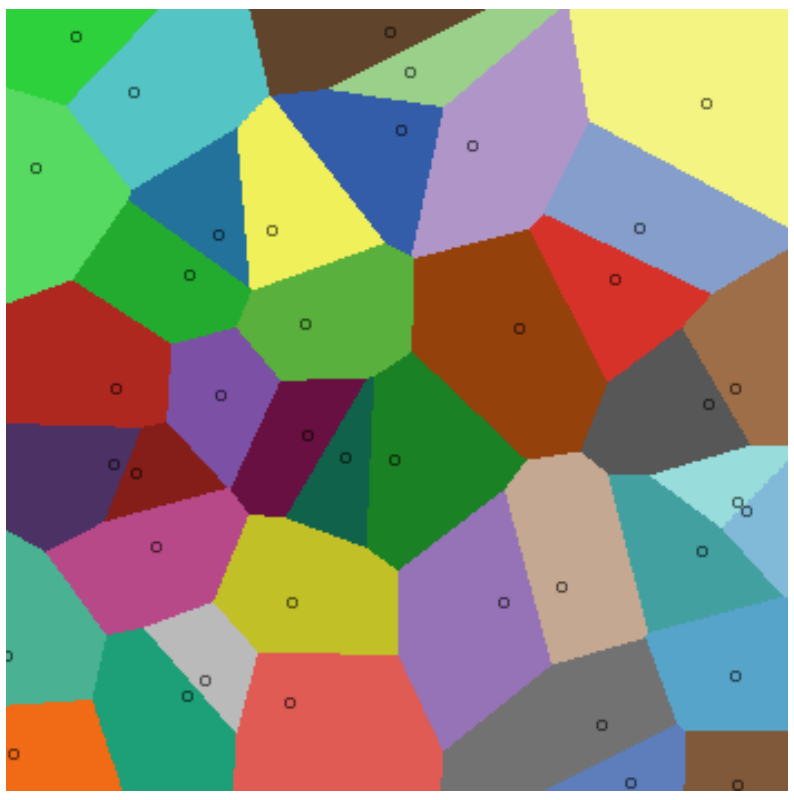

**Fig. 1.** Shows an example of a Euclidean space with generator points and their corresponding Voronoi polygons. The polygons are "emerged" from the collective positions of the generator points such that all points in each area corresponding to a polygon are closer to the corresponding generator point than to any other point [20].

We call the assumed behavior resulting from information symmetry *Voronoi* behavior or *Voronoi* movement. We propose this behavior using the spatial concept of a Voronoi polygon [19] as a base. A Voronoi polygon or diagram—as it is commonly known—is a partitioning of a spatial plane such that "all locations in the Voronoi polygon are closer to the generator point of that polygon than any other generator point…in Euclidian plane" [18]. In other words, it is the space such that maximum territory is created for each generator point without overlapping the area belonging to any other generator point. Figure 1. shows an illustration of generator points and their respective Voronoi polygons.

Voronoi movement essentially amounts to driver agents receiving location information about the nearest driver agent and moving away from them so as to increase the potential of picking up a new customer and reducing local competition—a diverging topological behavior where each agent maximizes the distance and the territory between self and all other agents. This viewpoint is a corollary to the Voronoi polygon—from the view of the generator point (agent) not the adjacent spatial points in the polygon, hence we call this behavior Voronoi behavior.

We compare Voronoi movement behavior with a random movement pattern where driver agents move randomly across our spatial grid until they are close enough to a rider agent to execute a pickup. The random movement behavior is a reflection of driver agents having no knowledge of where other drivers and where customers might be. In this iteration of the model, we assume agents do not learn.

In a high-fidelity model, it would likely be the case that drivers would learn about rider behaviors and adjust their own behaviors accordingly. However, as you will see in the results section, even with this simplification, our baseline model offers interesting conclusions nonetheless. Finally, rider (potential passenger) agents do not move in this iteration of the model but enter the simulation at a spatially random location on the Washington, D.C. geographic lattice. Figure 2 and 3 show the graphical representation of the model which was developed in NetLogo [20] and subsequent sections will discuss model design particulars.

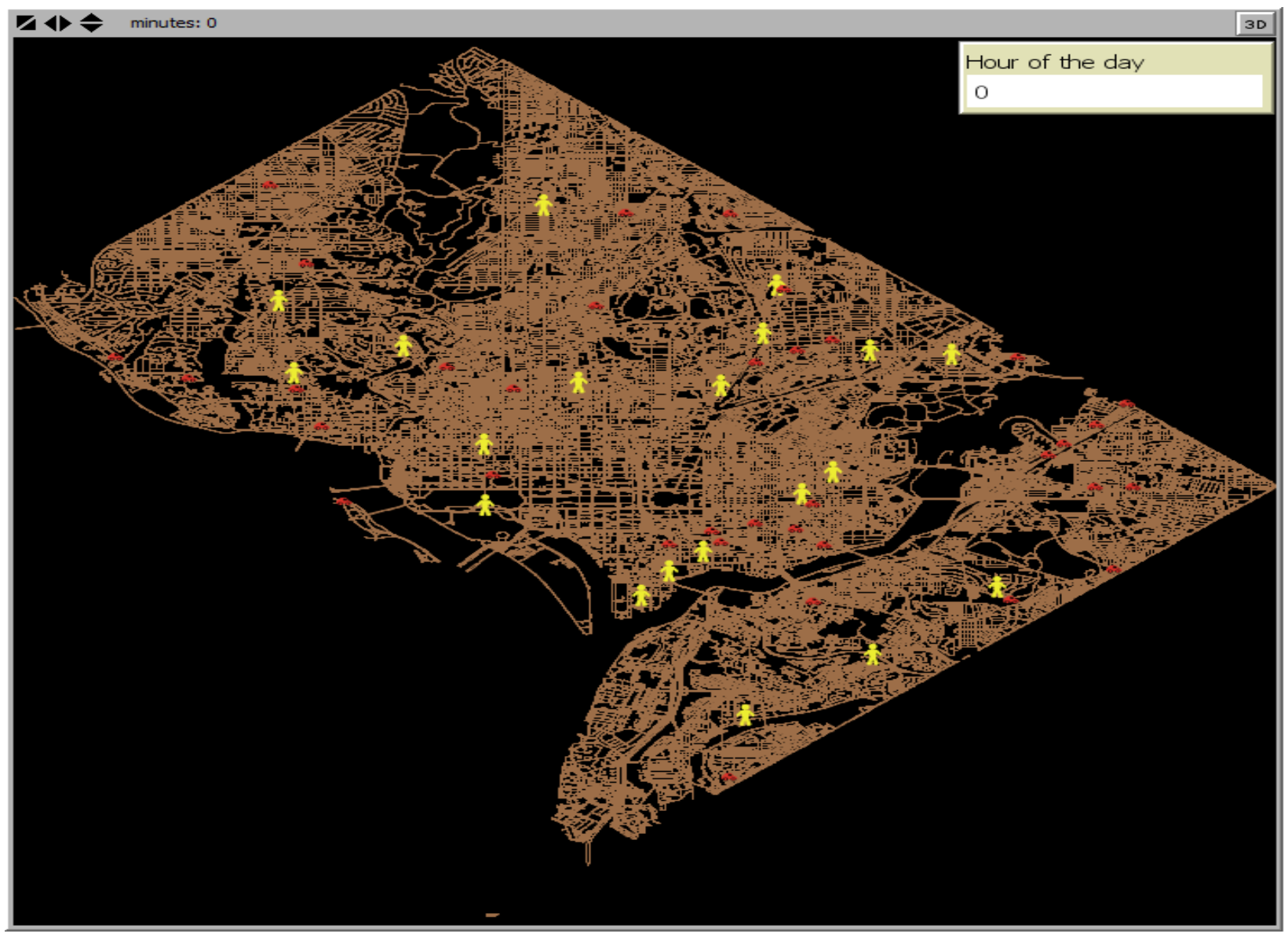

**Fig. 2.** This is the graphical representation of the model. Yellow agent types (person icons) are rider agents. Red (car icons) are driver agents. When a driver agent performs a pickup, their color turns from red to white to display that they are no longer available.

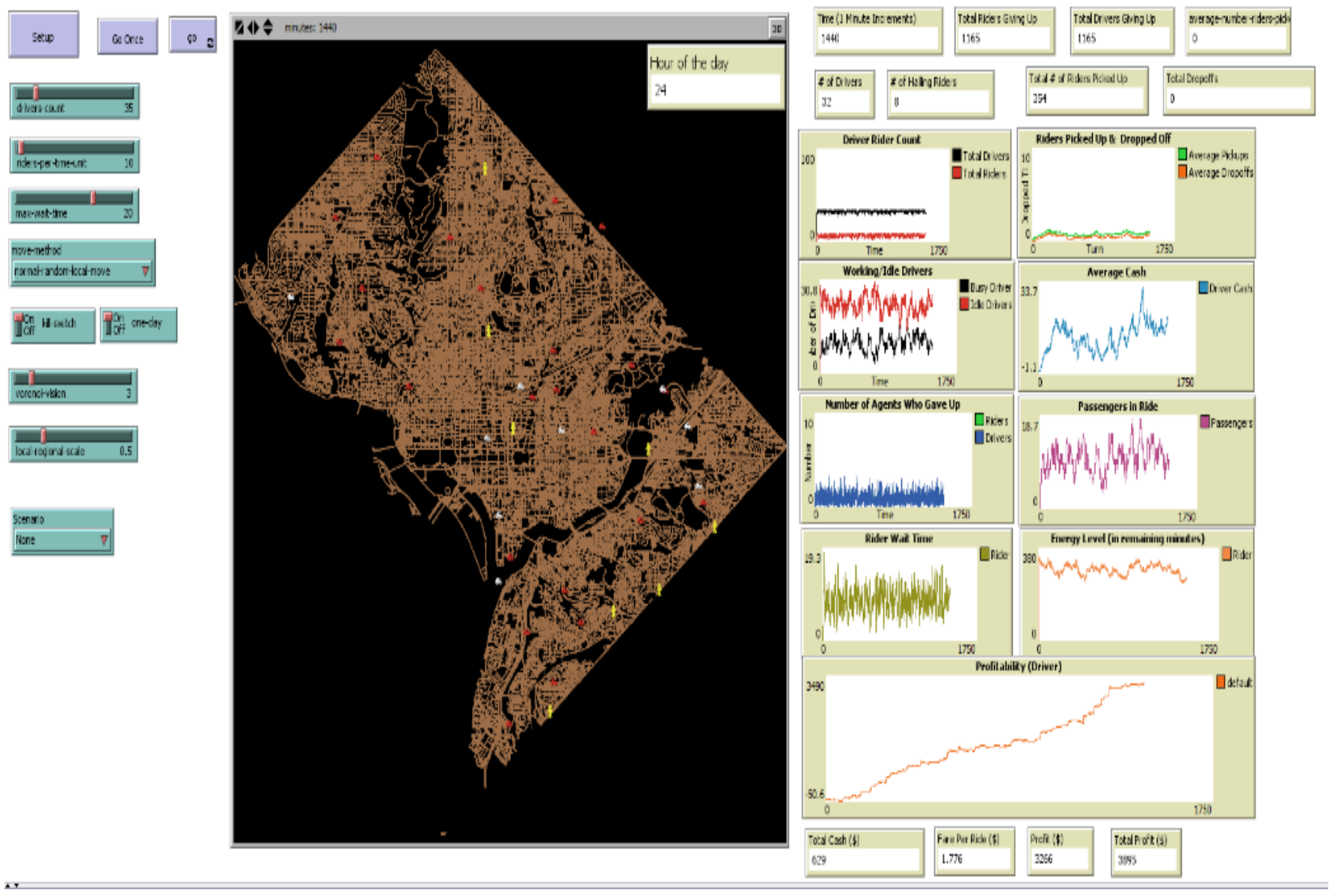

**Fig. 3.** This model was implemented in NetLogo 5.3 and utilized open access data from the Washington, D.C. government website. In this figure, we show the graphical user interface of the model. Inputs, such as the number of drivers and the number of potential riders active at any moment are complemented by outputs on the far right such as profitability, average cash on hand, and passenger pick-ups.

# *2.*Model & Methods

IAMWARM was implemented in NetLogo 5.3 and utilized the GIS extension native to the platform to import map and GIS data into the model. The model's spatial configuration was based on a road network imported from the Washington, D.C. government Open Data Project website[3], which included highly accurate, editable shapefiles. The data included feature labels for roads and intersections. These features are used in the instantiation and location initialization of agents, and for certain critical agent behaviors to be discussed in later sections of this paper.

At the current iteration of the model we chose not to include additional layers of geographic information for simplicity (only the road network was included), but in future iterations utilizing the Open Data project more broadly can be advantageous in increasing the efficacy of our model specifically by adding more spatial configuration data. Figure 4. provides a complete graphical summary of the model's logic.

The road network was imported to NetLogo and an internal spatially-equivalent configuration was assigned (labeling). All roads were labeled internally by a variable to help identify spatial cells that contained a road, versus spatial cells that did not. This would later be an important step when designing the movement choices of agents in the simulation since all movement and agent entry will occur on road cells as one might expect.

The model contains two agent types: drivers and riders. Drivers can move across the model space, but only on cells that contain a road, while riders do not move, but can only be initialized on roads, specifically intersections.

The movement of the driver agents was designed to be based on a direct line of sight—that is—although the drivers must always remain on roads, we assume that following actual traffic routes would not provide a negligible difference in destination arrival times. This is mainly due to the size of our spatial lattice which numbers in the several thousand. Moreover, for our research goals, it suffices that driver agents move in a direct path to their destinations once rider agents are picked up. In future iterations of the model, traffic and road direction movement could be taken into account to create a greater sense of realism. For now, driver agents move on roads in a direct fashion to their destinations.

## *2.1* Agent Behaviors

Before we discuss the specifics of agents used in our model we define the terminology used in the model. We define driver agents as those agents who are intending to pick up a rider. A rider agent is an agent who has been instantiated and can be picked up by a driver. Once a driver agent picks up a rider agent, the rider agent becomes a passenger. Passengers are not agents and do not interact with their environment. In other words, riders who become passengers simply become an attribute of the driver agents, releasing with them certain data points which are then used by the driver agents post pick-up. This terminology will be used throughout the model description.

[3] www.dcogc.org

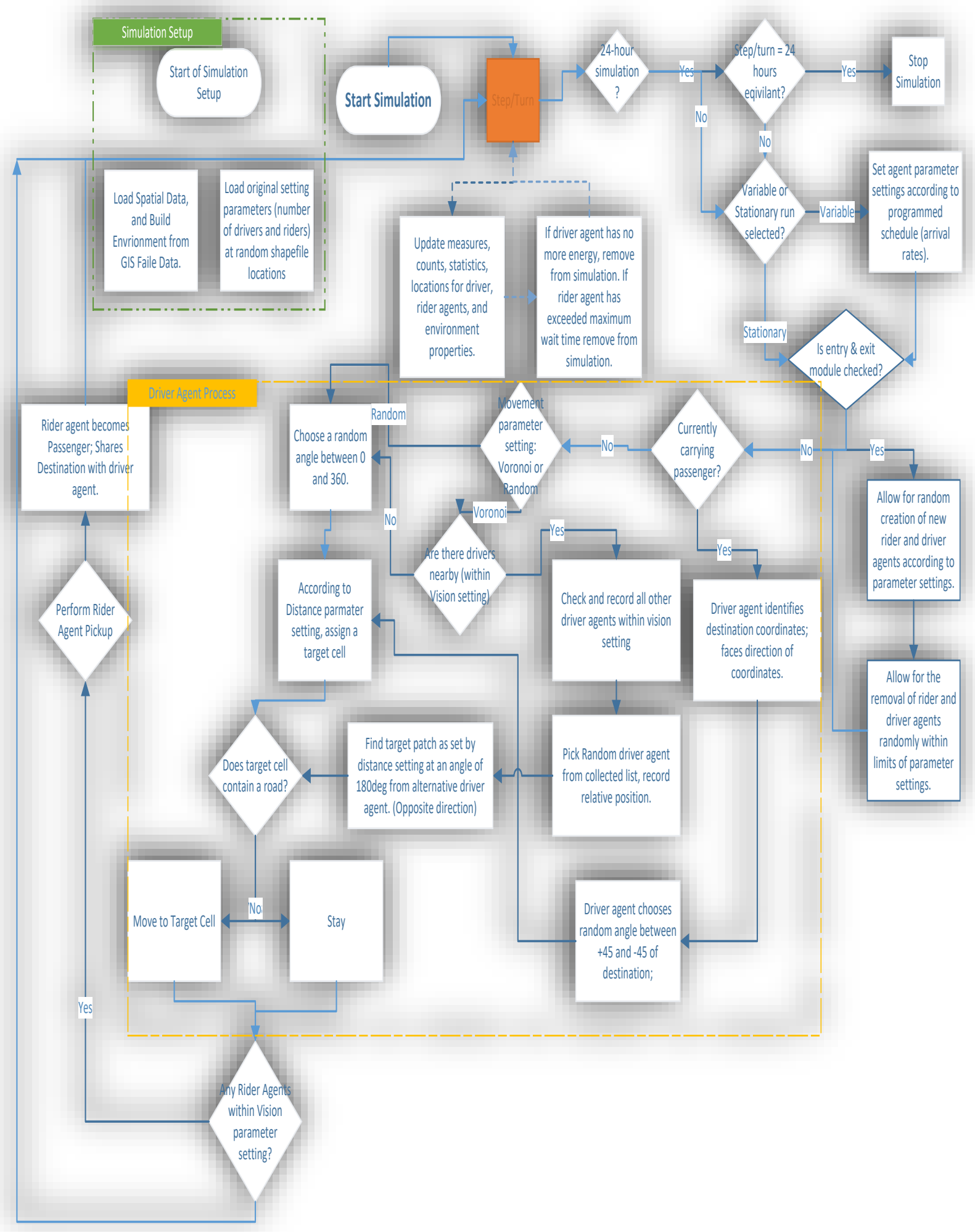

**Fig. 4.** A complete breakdown of the model's process is presented in this figure. The area labeled inside the yellow box is where the majority of driver agent behavior takes place. The small orange box at the top of the diagram represents the model moving to the next time unit/time cycle. Note that the majority of rider agent interactions are included as part of the driver agent behaviors since much of the rider agent behaviors in the model is restricted to entering or leaving the simulation.

### *2.2* Driver Agents

Driver agents are instantiated in the initial setup of the model according to a user input parameter and a random distribution set to uniform properties. Spatially, the driver agents are initialized at random road intersections of the geospatial model, and according to a randomization test score that allows for a greater randomized spatial distribution of agent objects. Driver agents are also instantiated throughout the model run according to the aforementioned user set parameter. Those agents are also instantiated at road intersections and their rates of arrival are also set according to a user set parameter.

Driver agents move from one spatial cell to another by first checking whether a road exists in the cell ahead. If it does, then an agent may move to that cell. If no road exists in that cell then for the random movement behavior without an active passenger procedure, agents are instructed to rotate a random number of degrees between 0 and 360 and repeat the process.

For driver agents who are carrying a passenger, the process is the same. However, the randomized degree value is set to be between -45 and +45 degrees if a road is not found directly in the heading of the driver agent. By doing so, we ensure that drivers are constantly moving in the direction of their destination, but are still able to overcome the majority of obstacles in their way, such as the lack of available roads to travel on. This method does have grounds in reality in that drivers who may not necessarily know precisely how to optimize their routes, may in general, pick a random route that they know to be in the general direction of their destination. In future iterations of the model a more advanced pathfinding algorithm could be adopted such as the A* pathfinding algorithm, but for our purposes, we assume that the difference is negligible and it is trivial to show so through a model run.[4]

Driver agents are assigned a number of attributes at instantiation and some are assigned as the model is run situationally. Attributes include energy level, cash-on-hand, time driven, current driver destination (if carrying a passenger), riders who are nearby, current passenger id, time the current passenger has been on a trip, how many riders the driver has picked up, and how many passengers the driver has dropped off, as well as a Boolean passenger variable indicating if the driver agent is currently carrying a passenger. The attributes are more critical to the progress of the model at varying times through simulation runs, depending on the active phase of the drive-pickup-drop-off-drive cycle of the driver agents. Therefore, a deeper explanation of the attributes and their relevance is appropriate at this time.

We assign a level of energy to every driver agent set to be a random number following a normal distribution between a 4 and 8-hour range translated into model time units (which is set to be at 1 minute per time unit) by estimation. The underlying assumption is that the majority of drivers will only be able to drive, regardless of their level of success, for a period determined by time availability and physical and/or mental fatigue. This is an appropriate assumption and the Gaussian shape of this distribution is not far-fetched. Time-driven is a variable that counts the amount of time driven by each driver and is used to display and calculate the model's summary statistics.

[4] For a video of a typical model run, please visit https://www.youtube.com/watch?v=apJEvDl4aqc

Cash-on-hand is the variable attribute that stores the accumulation of fares each driver agent has gained, as well as the variable in which cash is deducted (vehicle and transportation costs) for drivers who are not carrying active passengers. In other words, it is the driver agents' total utility and financial gain at any given time-period.

Pick-up count and drop-off count are variables that store the total number of riders successfully picked up from their initialized location and successfully dropped off at their destination, respectively. The Boolean passenger variable shows whether the current driver is currently seeking a rider or already has a passenger (as discussed earlier, riders that receive a pickup by a driver agent become 'passengers'—simply an attribute of the driver agent) and is used in a number of important model mechanics. Passenger-id is a variable that stores the id of the rider currently within the driver agent's vehicle when she becomes a passenger. It is equivalent to the driver asking for the rider's name prior to pick-up and is used in the verification process of the model to ensure that drivers are successfully picking up intended riders.

Variables for nearby riders and nearby drivers were also implemented as attributes of the driver agents. The first being the number, and id of any nearby riders waiting for pickup. This attribute is used to evaluate whether there are any riders nearby available for pickup. The second is whether there are any nearby driver agents, and is used in the Voronoi movement mechanism. The details of the Voronoi movement mechanism will be discussed in a subsequent section.

### *2.3* Rider Agents

Rider agents are instantiated at initialization of the model at random locations (intersections). Rider agents do not move but await their intended pick up in the same location. This is in line with expected behaviors of ridesharing service customers. Rider agents are instantiated utilizing a user-set input and a random variable to allow for some stochasticity in the model runs. Spatially, and in a similar fashion to driver agents, they are placed randomly across the available intersections of the model's geographic configuration, and only on road cells.

Rider agents are instantiated throughout the model's runs but at a rate per time unit user fixed parameter, unlike driver agents who typically remain in the model's space until they decide to *leave* (randomly) or because they have exhausted their energy variable and are replaced stochastically up to a maximum user set parameter. In other words, while the number of driver agents is set by a maximum capacity global (exogenous) variable, the number of rider agents is set to be a rate of arrival following a probability distribution. For simplicity, we chose the arrival rate probability distribution to be normal, though in subsequent iterations of the model testing of other probability distribution types, such an exponential arrival function, would be necessary.

Rider agents possess two attributes for this iteration of the model. The first is the rider destination, which is a randomly assigned destination converted to the spatial coordinate equivalent. This destination is assigned at the moment of instantiation of the rider agents. The second is the wait time variable which is a count of how long a rider has been waiting for pickup and which is used to compare to a user-set input to determine whether the rider agent should look for 'alternative' transportation methods (like metro or bus service). A user designated input allows for varying the waiting time of rider agents. For typical model runs we assigned this variable to 20 minutes. Once a rider agent reaches their maximum waiting time assigned they leave the simulation. This is a proxy behavior for the rider agent attempting to find alternative modes of transportation to their destination.

## *2.4* Model Mechanics

The model relies on user inputs for the number of drivers (capacity), the number of riders arriving per given time unit, the maximum waiting time for rider agents, whether or not to use random movement while attempting to find a rider or whether to maximize distance from any other drivers in the area of assigned vision (Voronoi movement). Based on these inputs and the parameters of the model a typical model run behaves as follows.

Rider and driver agents are instantiated on a highly accurate road map of Washington, D.C. at random intersections in accordance with the input parameters assigned by the model user. Once the model is run, rider agents are spawned while others, according to the user input, will leave the simulation. The same is applied to driver agents.

Driver agents move according to one of the predetermined movement methods (random movement or Voronoi movement based on which version of the simulation is run) until they are within a user-set proximity setting of a rider agent (for our model the vision was set to 3 cells). During this time, driver agents lose energy at a rate of 0.75 per time unit and lose cash at a rate of $0.1 per time unit.[5]

Once a driver agent is within a designated (by modeler) vicinity of the rider agent, a pick-up occurs, and the rider is converted to a passenger. A transfer of the destination of the rider occurs from rider agent to the driver agent at this time. Once a driver agent has successfully executed a pickup, the agent can no longer make any additional pickups, and it is at this point that the driver agent begins to accumulate ‘cash’, set at a rate of *$2.00 + 0.60/*time unit. Calibration of the model was conducted to reach a dollar amount that could be probable through qualitative observations of distance and time of real-world trips versus the model’s spatial geometry. The key was to set the fare rate to include a fixed amount and a variable amount so as to reflect actual ridesharing services.

---

[5] We assigned these cash variables based on a rough estimate of distance travelled versus fare/ride gained from observations and experiences with a ridesharing service.

Driver agents then proceed in a direct path to the coordinates of the transferred destination while earning ‘cash’ and losing ‘energy’. Once they reach their proximate destination, a drop-off is executed, their passenger-carry variable is reduced from 1 to 0, and a successful trip is recorded as being now completed. All relevant attribute and model-level variables are updated with this new information. The driver then continues to move searching for new riders and repeats the drive-pickup-drop-off process.

Rider agents who are not picked up within their waiting time-period limit find alternative transportation and leave the simulation, while driver agents who are not carrying a passenger could “give up” and leave the simulation. The latter could also run out of energy and leave the simulation due to fatigue. Typical runs are for a 24-hour period, but a model user can run the model indefinitely if they desire.

### *2.5* Model Inputs and Parameters

We implemented 7 inputs in our model that can be assigned and varied by the user. Table 1. summarizes those inputs and contains their descriptions. The most important of which are the maximum capacity for driver agents (drivers-count) and the rate of entry of new rider agents into the model (riders-per-time-unit). Other inputs are also critical but were not tested in a significant way—though adjusted for calibration and realism. Those are the Voronoi vision setting which controls how far driver agents can see other driver agents, the local-regional scale which amounts to an adjustment for the speed of movement of the driver agents, and a binary-switch which turns on or off the possibility for both driver and rider agents leaving the simulation randomly.

In Table 2, we describe the parameters and distributions used in various parts of the model. As mentioned in an earlier section we used a normal distribution of varying means and standard deviations as the basis for a number of statistical tests and parameter values so as not to add any additional unverified assumptions or complexity to the model. Hence, we rely on the Central Limit Theorem heavily. However, in future iterations of this model specific testing and data collection of the distributions’ parameters must be undertaken and compared to real data from a ridesharing service.

**Table 1. Inputs of the model with typical value assignments.**

| Type | Input | Description | Typical Value(s) |
|---|---|---|---|
| Driver Agents | Driver Agent Number | Assigns a maximum number of drivers to be active at any given time unit | 50-150 |
| | Normal Random Movement (Choice) | Sets the movement behavior of drivers to be of a random nature while they await a rider pickup | N/A |
| | Voronoi Movement (Choice) | Sets the movement behavior of driver agents to follow a Voronoi-distance maximizing method | N/A |
| | Voronoi Vision | If Voronoi Movement is chosen, sets the Voronoi movement vision distance | 3 |
| | Local-Regional Scale | Sets the vision and movement range for driver agents. This amounts to a speed setting and is used to calibrate the model. | 0.5 |
| Rider Agents | Riders Active Per Time Unit | Sets the rate by which new riders enter the simulation and await pickup | 20-75 |
| Environment | Scenario (Choice) | Sets the model into a run type where an expected rate of arrival for riders and an expected maximum capacity for drivers is set at different hours of the day. | Saturday |

**Table 2. Parameters used in model mechanics.**

| Parameter | Type | Value | Description |
|---|---|---|---|
| Driver Agent Placement | Test X > 0.5 | Normal (1,1) | Tests whether a random number from a normal distribution with a mean of 1 and standard deviation of 1 is greater than 0.5. If so, placement of a driver agent succeeds at a given intersection. |
| Rider Agent Placement | Test X > 0.5 | Normal (1,1) | Tests whether a random number from a normal distribution with a mean of 1 and standard deviation of 1 is greater than 0.5. If so, placement of a rider agent succeeds at a given intersection. |
| Driver Agent Energy | Attribute | Normal(360,120) | Sets the energy of a driver agent at instantiation as a number drawn from a random distribution with a mean of 360 and a standard deviation of 120 (minutes) |
| Kill Count | Variable | \|(Normal (0,1)\| | Sets the number of driver and rider agents who will leave the simulation, randomly without depleting their energy (driver agents) or reaching maximum wait time (rider agents) to be the absolute value of a random number drawn from a normal distribution with mean 0 and standard deviation of 1. |

## *2.6* Model Outputs

A number of outputs were included in the model to assist in the verification process, to understand model mechanics and to derive results from model runs. Table 3. lists those outputs and their descriptions. Our focus was to understand driver agent utility given some set of inputs, parameters, and pre-conditions. There are many forms of driver agent utility to consider, each of which would require a focus on a different set of output measures. For this iteration of the model, we chose to focus our attention on total driver agent utility in the form of total profit from each model run. We include no outputs to measure ridesharing service utility or rider/passenger utility in our final analysis and conclusions, however, a number of outputs aimed towards the measurement of rider agents, passengers, and ridesharing service utility are designed into our model and are displayed to the user. We hope to expand on our analysis of system utility by considering rider, passenger and service utility in future iterations of the model.

**Table 3. Model outputs and measures**

| Output | Description |
|---|---|
| Number of driver agents active | The number of driver agents active in the model |
| Number of rider agents active | The number of rider agents active in the model |
| Total riders giving up | The total riders giving up based on randomly set parameters |
| Average number of riders picked up per time unit | The average number of riders picked up per time is calculated for each time unit and displayed |
| Total number of rider agents picked up | This is the total number of rider agents converted to passengers |
| Total number of successful drop-offs | Total number of successful drop-offs, which tends to be lower than the number of pickups as some driver agents don't reach their destinations |
| Number of idle driver agents | Number of driver agents without an active passenger |
| Number of working driver agents | Number of driver agents with an active passenger |
| Average cash on hand | The total amount of cash held by all driver agents (active) |
| Number of agents who left (randomly) | Number of agents who left the simulation due to random tests |
| Number of passengers in active trips | Number of passengers carried by driver agents |
| Average wait time | Rider agent average wait time |
| Average energy level | Driver agent average energy level |
| Total cash with active driver agents | Total cash for all active driver agents at any given time unit |
| Average fare per ride | Average fare per ride at any given time unit |
| Total profit generated | Total profit generated by all activity of the model |

## *2.7* Scenarios

Much of the model can be run in a stationary mode—that is—it can be run in a form of equilibrium where driver agents and rider agents arrive at predetermined rates resulting in a constantly changing but variably fixed dynamic. This is interesting for general runs, verification of model mechanics and quality, as well as to gain a general understanding of the ridesharing process. It is trivial to hypothesize that in any transportation system the rates by which riders and drivers arrive, interact, and exit are variable but also subservient to the city (the spatial lattice) in which the ridesharing service operates. This would include seasonal variables such as the time of year, month, day, and time of day. Additionally, rider and driver rates and activity are also affected by current events, traffic, roadworks, weather patterns and other exogenous factors. Therefore, though running the model *in situ* yields important insights, it is important to run experimentation in some variable scenario for comparative reasons and for a closer approximation of real-world dynamics simply because the variance itself could yield insight.

Therefore, based on anecdotal evidence gained from the author's registering with a ridesharing service and gaining first-hand experience in typical driver decisions made, we develop a scenario which is not entirely hypothetical in order to test the model's effectiveness under varying conditions. In future iterations of this model, we intend to develop scenarios grounded in real data collected and to develop a number of them to test different scenarios under different conditions without such heavy reliance on qualitative observations.

For this model iteration, we conducted a test of one scenario—the "Saturday" scenario which varies only arrival rates of both driver and rider agents according to what might be expected on a typical weekend day—Saturday. Table 4. provides a summary of the scenario and reasoning, where appropriate, for selection of scenario inputs and parameters. As you will see from the table, we varied the arrival rates and expected capacities of rider agents and driver agents respectively. For example, on 'Saturday' we would expect high customer demand for the hour before 'lunch' as many city dwellers may be engaging in social activities in the subsequent hour and so they intend to arrive before 'lunch-hour'. In this time-period (11 AM) and in this specific scenario, we expect that in anticipation of higher customer demand the number of driver agents may also increase, and so we increase the maximum capacity of the driver agents.

**Table 4. This is a variable model scenario used in testing the model under realistic conditions drawn from anecdotal observations.**

| Hour | Driver Capacity | Rider Rate of Entry | Reasoning/Explanation |
|---|---|---|---|
| 5AM | 10 | 5 | Early Morning - Airport Traffic - Mostly Quiet |
| 8AM | 20 | 10 | |
| 11AM | 50 | 20 | Drivers starting their day for the Saturday Brunch/Lunch |
| 12PM | 5 | 10 | Lunch Time - low activity |
| 1.30PM | 45 | 25 | Post-Lunch Rush |
| 2.30PM | 50 | 15 | Post-Lunch Rush |
| 3.30PM | 25 | 10 | Stationary Activity |
| 5.30PM | 40 | 5 | Evening Drivers Beginning Their Shifts |
| 6.30PM | 45 | 5 | Evening Drivers Beginning Their Shifts |
| 7.30PM | 60 | 30 | Night Activity Period - Riders are going out to social events |
| 8.30PM | 80 | 40 | Night Activity Period - Riders are going out to social events |
| 9.30PM | 100 | 40 | Night Activity Period - Riders are going out to social events |
| 10.30PM | 90 | 10 | Low Rider Activity - Riders are at their destinations. Drivers still on the road expecting a rush of new riders. |
| 11.30PM | 80 | 10 | Some drivers give up, exit the simulation |
| 12.30AM | 75 | 30 | More drivers give up. Riders beginning to end their work shifts. |
| 2AM | 65 | 30 | More drivers give up. Riders beginning to end their work shifts. |
| 4AM | 35 | 10 | End of night traffic. End of 24 our cycle. |

## *2.8* Testing, Verification, and Validation

To test whether spatial movement behaviors can affect driver financial gain, we aimed to compare Voronoi movement prior to agent-pickup with random movement prior to agent-pickup. In other words, having drivers access information about other local drivers and having chosen to 'spread out' maximizing their personal territory and likelihood of rider pickup, when compared with random movement, regardless of where other driver agents may be. This collection of tests translates to:

1. Comparing Voronoi movement with random movement under stationary conditions (constant capacity and arrival times through entire run)
2. Comparing Voronoi movement with random movement under variable conditions, namely a 'Saturday' scenario (varying arrival rates and capacity for rider agents' entry and driver agent entry).

Therefore, to test our model we conducted 4 standard runs: A scenario-based set of runs with a comparison of random movement and Voronoi movement, and a stationary standard run with Voronoi movement and random movement. We also conducted a number of verification and validation tests to ensure that the model is run correctly as well as that it is running as intended. We summarize those efforts in Table 5.

**Table 5. Verification and Validation Methods**

| Goal | Method | Result |
|---|---|---|
| Verify that road network imported correctly | Display and check spatial cell attributes | Success |
| Verify that agents instantiate on roads, specifically intersections. | Compared spatial coordinates of agents with cell coordinates of intended intersection | Success |
| Verify that agents instantiate with the correct attribute values | Displayed agent attributes at random | Success |
| Verify that the movement of agents is as intended | Visual observation and numerous model runs | Success |
| Verify that the model mechanics for rider entry functions as intended | Raised and lowered the rider entry value and monitored expected increases or decreases in outputs | Success |
| Verify that the rider capacity functions as intended | Raised and lowered the rider entry value and monitored expected increases or decreases in outputs | Success |
| Validate Pickup Mechanics | Verified through the transfer of rider id and destination | Success |
| Validate Drop-off Mechanics | Observed attribute changes for increases in drop-off values | Success |
| Validate agent attribute changes through model runs | Through many runs and observations | Success |

# 3. Results

## *3.1* A Typical Run

We present typical model results for the 4 possible variations of our test runs. Figure 5-7 and table 6-7 show plots and summary results of our model run. Figure 5-7 contain the total and accumulated financial gain of all agents with total profit on the y-axis and time units on the x-axis. This measure includes the financial gain made by agents who have left the simulation due to fatigue or for any other reason.

We see no real and substantial difference in terms of total profit (financial gain) between either of our stationary conditions model runs for random movement and Voronoi movement (yellow and grey). That is—whether drivers chose to 'spread out' or move about randomly in the hopes of picking up more customers did not affect, on average, their financial gain. In fact, the difference was comfortably within 2 standard deviations for both Voronoi and random movement runs.

Remarkably, for the variable scenario runs a stark difference emerged between the two spatial movement types, unexpectedly. Divergence in the profitability between random movement choices and Voronoi movement choices for driver agents was clear, and exhibited in both the total profit made by driver agents in a 24-hour run (Figure. 6-7) and in the summary statistics of the model run as a whole (Table. 7).

Specifically, we can comfortably note that Voronoi movement for driver agents provides greater utility (financial gains) for drivers when varying rates are executed on the agents' arrival rates i.e. when a scenario is utilized. Where for stationary model runs neither movement method prior to rider pickup provided any visible change in driver agent utility. Consequently, our observations and analysis of the model run took a focus on the variable scenario runs, and more precisely on the moments of divergence of the variable scenario random movement run when compared to the Voronoi movement variable scenario model run.

**Table 6. Summary statistics for stationary run for both random and Voronoi movement**

| Stationary Random | | Stationary Voronoi | |
|---|---|---|---|
| Mean | $ 6,284.02 | Mean | $ 6,483.38 |
| Standard Error | $ 94.35 | Standard Error | $ 88.71 |
| Median | $ 6,161.00 | Median | $ 6,400.90 |
| Standard Deviation | $ 3,580.49 | Standard Deviation | $ 3,366.40 |
| Maximum | $ 12,426.55 | Maximum | $12,059.45 |

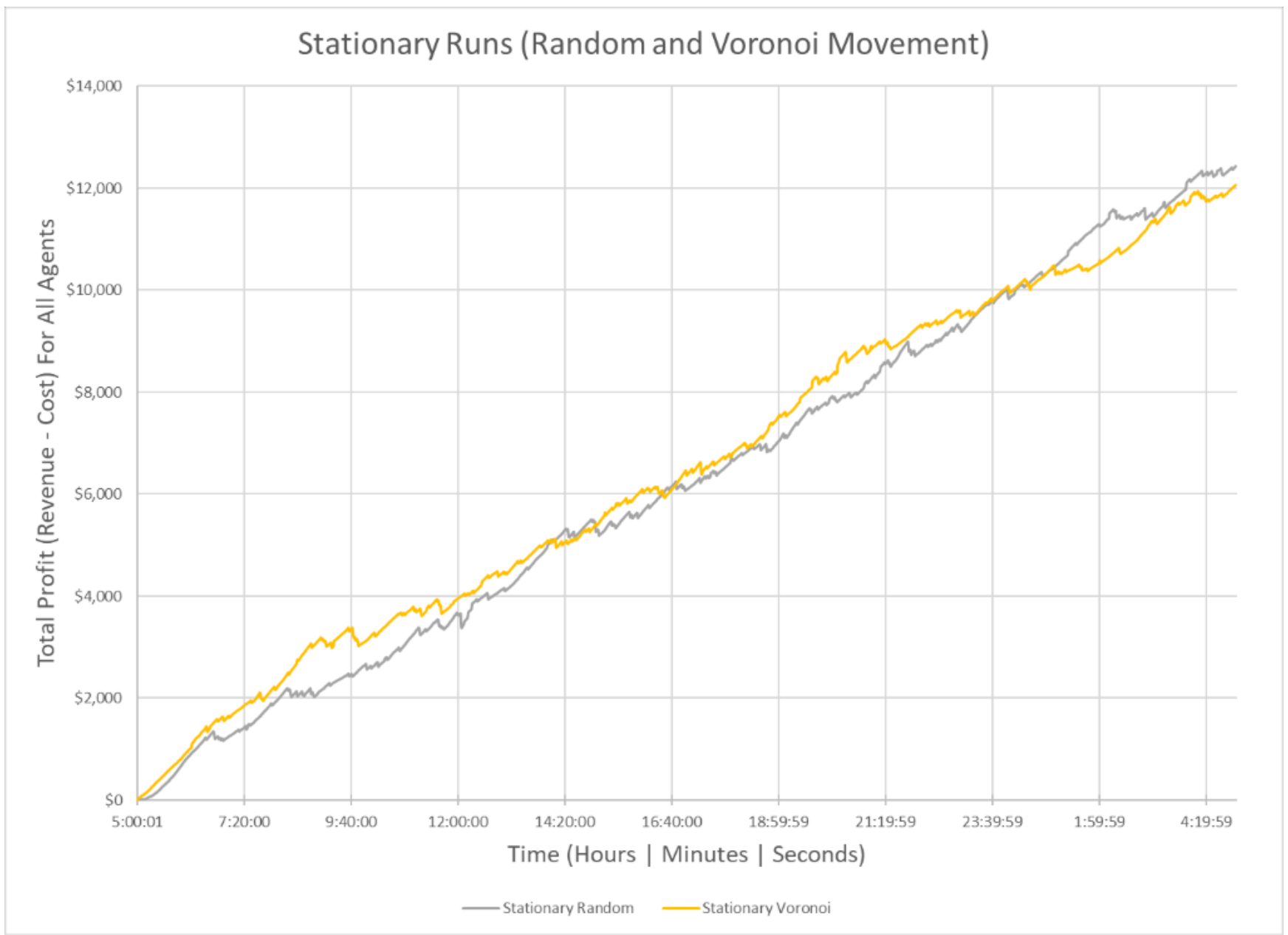

**Fig. 5.** This figure shows the total profit made by all driver agents over a 24-hour period—a single run of the model for both random driver movement (gray) and Voronoi driver movement (yellow). The figure shows that while there are stochastic gains made at different times in the model run by both random and Voronoi movement behavior, there are no clear advantages in utilizing either behavior type when the arrival conditions of drivers and riders are stationary.

**Table 7. Summary statistics for variable properties run for random and Voronoi movement**

| Variable (Saturday) Random | | Variable (Saturday) Voronoi | |
|---|---|---|---|
| Mean | $ 2,845.05 | Mean | $ 3,931.65 |
| Standard Error | $ 68.96 | Standard Error | $ 87.31 |
| Median | $ 1,828.90 | Median | $ 3,332.50 |
| Standard Deviation | $ 2,617.00 | Standard Deviation | $ 3,313.05 |
| Maximum | $ 8,195.65 | Maximum | $ 10,565.45 |

More importantly than the observation that Voronoi movement outperformed random movement only in variable run conditions, is whether we can deduce precisely where the divergence between the two behaviors began to take shape under variable entry conditions.

The first instance of divergence in a typical run occurs at around 1 PM (marked on Figure 7) into the model run which we hypothesize as being representative of "lunchtime" activity and model as being a constant driver agent capacity of *5* drivers and an arrival rate of new riders of 10. Both rates are a reduction from the 11 AM hour which had a maximum capacity of 50 driver agents and 20 respectively. At 1.30 PM the capacity for new driver entry increased to 45 and the rate of arrival of rider agents also increases to 25 (Table. 8). Figure 7 has both the random movement and Voronoi movement drawn with separate y-axis on the same time-scale (x-axis) so as to allow us a better visual comparison of both run-types, and we can see that at this lunch-time hour a slight divergence of performance begins to emerge, allowing drivers who are using Voronoi movement to make placement decisions that outperform those that move randomly.

**Table 8.** This table shows the relevant arrival rates for drivers (left) and riders(right) for the first point of divergence in the variable conditions model run.

| | | | |
|---|---|---|---|
| 11AM | 50 | 20 | Drivers starting their day for the Saturday Brunch/Lunch |
| 12PM | 5 | 10 | Lunch Time - low activity |
| 1.30PM | 45 | 25 | Post-Lunch Rush |

The most salient divergence between the performance of the two behaviors we tested occurred at around the 3.30 PM time-period. Figure 7 shows the stark difference in performance and thus in utility-gain between the two behaviors. Our scenario at this time-period calls for the decrease of both driver capacity and rider entry from the 2.30 PM period (from 50, 15 to 25, 10, for driver and rider agents respectively). Table 9 summarizes the relevant part of the scenario run.

**Table 9.** This table shows the relevant arrival rates for drivers and riders (left, right) for the second point of divergence in performance of the variable conditions model run.

| | | | |
|---|---|---|---|
| 2.30PM | 50 | 15 | Post-Lunch Rush |
| 3.30PM | 25 | 10 | Stationary Activity |
| 5.30PM | 40 | 5 | Evening Drivers Beginning Their Shifts |

Thus, we saw a divergence in performance for a case when both driver and rider entry were increasing simultaneously, and for a case where they were decreasing simultaneously as well, which dispels any notion that divergence in performance would solely be due to a decreasing rate of one arrival rate while another was increasing. We will propose candidate theories in the discussion section of this paper.

The objective of our "lunchtime" change in both driver and rider agent demand and supply was to create drastic changes similar to that what would be expected in a major metropolitan area during this time period. What is critical to note is that this performance difference—this emergent pattern—is only seen under variable run conditions, and not stationary run conditions. Consequently, during a statistical analysis for our model's stationary run, we find that the mean, median and maximum financial gain (by all agents) during a 24-hour period was not significantly different between driver agents employing a Voronoi movement versus random movement behavior. Wherein the variable ("Saturday") scenario run, the median, mean, and maximum were contrasted, with Voronoi movement outperforming random movement decisions on the aggregate; though the majority of the performance improvements came from the time-periods where Voronoi movement allowed a greater rate of customer pick-ups (1 PM and 3.30 PM). It was not immediately apparent that Voronoi movement outperformed random movement for every time-period of the variable run scenario.

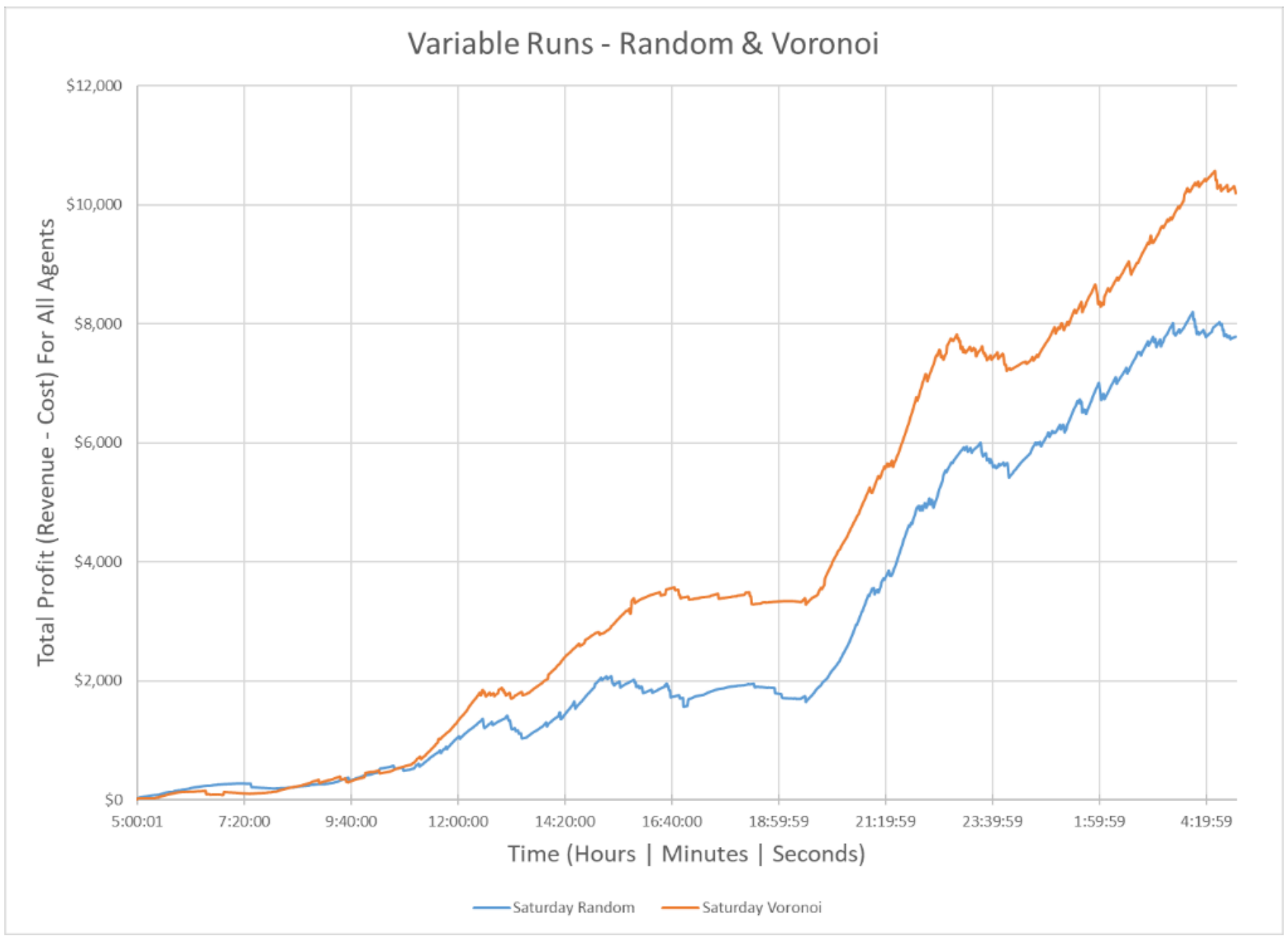

**Fig. 6.** This figure shows the total profit made by all agents over time. Not the divergence between Voronoi movement (Orange), and the random movement (Blue) under variable conditions. Voronoi movement outperforms random movement.

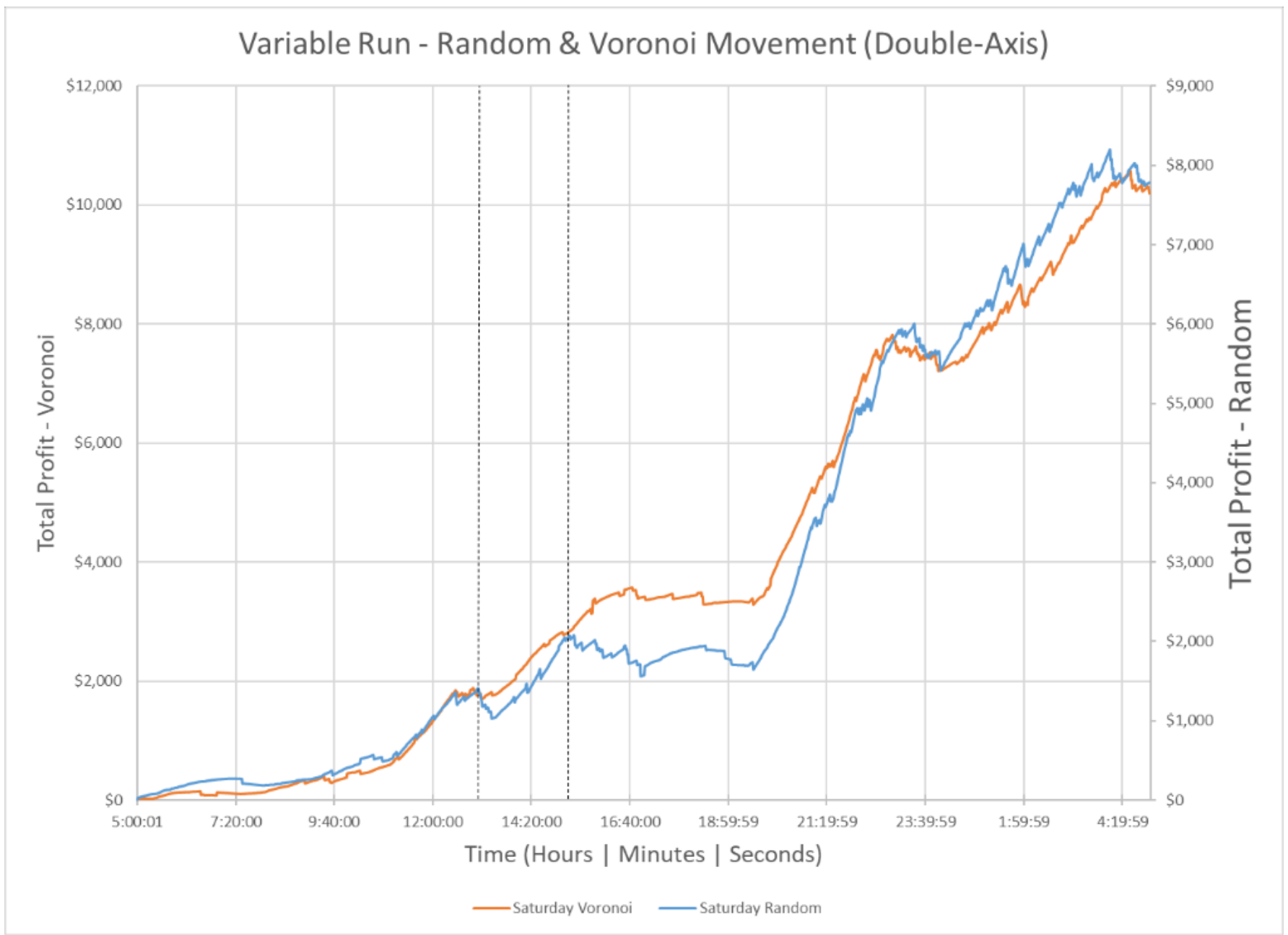

**Fig. 7.** This figure utilizes 2 axes, one for the variable conditions scenario run (Saturday) Voronoi movement (Left axis) and another for the random movement (right axis). This makes for easier visual comparison and yields insight into precisely which point in time the divergence between the two behaviors would typically occur. In this particular run it is clear that Voronoi movement begins to outperform random movement at 630 minutes.

# *4.* Discussion

Clearly, the model shows that driver agent positioning prior to rider agent pickup influences the financial utility gained by driver agents (consequently this also means that riders receive more consistent pickups with shorter wait-times). But this emergent phenomenon is not recognizable unless a realistic variable arrival rate scenario is utilized. I argue that this is true because of the *spreadin*g nature of Voronoi movement. In highly volatile rider-supply and driver-demand areas of our variable scenario runs, spreading-out behavior ensures that drivers are more evenly distributed, and by being so they are better positioned to "catch" riders in a moment of higher demand. Where, if driver agents choose to continue moving and placing themselves randomly in moments of drastic change to supply and demand, their *catching* behavior is set to be limited and thus are unable to maximize their financial gain and adapt to their surroundings. This is the case for when both rates of entry for drivers and riders are increasing and decreasing simultaneously, thus it should be noted that opposite signs for the first derivative of the profit variable are not a requirement for this phenomenon to occur.

The details of when this occurred are also important. In the "Saturday" scenarios, Voronoi movement did not outperform random movement in all variance combinations. For example, Table 10 shows a portion of the variable scenario at the 10.30 PM time-period. Note the decrease in driver capacity throughout the listed time-periods from 90 to 75, while rider arrival rates remained constant and then increased to 30 per time unit. Figure 7 shows that in this time-period there was no divergence in performance between Voronoi and random movement, even though elements of both the rise and fall seen in the scenario portions (where visible drastic change was present) was also embedded in this particular sequence of agent arrivals.

**Table 10.** A portion of the "Saturday" scenario that contains elements of the scenario portions where performance divergence was seen, yet, no divergence emerged for this sequence.

| | | | |
|---|---|---|---|
| 10.30PM | 90 | 10 | Low Rider Activity - Riders are at their destinations. Drivers still on the road expecting a rush of new riders. |
| 11.30PM | 80 | 10 | Some drivers give up, exit the simulation |
| 12.30AM | 75 | 30 | More drivers give up. Riders beginning to end their work shifts. |

Thus, we must then conclude that sudden increase and decrease in rider demand in conjunction with a steady or a slightly decreasing capacity (supply) provides a sudden spatial vacuum in the model's geographic configuration which is best enclosed by agents who are actively trying to move away from each other – a Voronoi movement pattern. But, that this pattern—that allows drivers to cover more space and 'catch' more riders—occurs only where there is enough drivers on the spatial geography such that there are actual additional riders who will be picked up by this movement. In other words, if decreasing or increasing variations in both agents' entry are occurring, Voronoi movement will provide drivers with an advantage over random movement, given that there are enough free drivers (without a passenger) and enough riders (without a driver) ready for pickup; an opportunity for maximizing utility must exist. After all, an increase in the spatial spread of driver agents allows for an increased probability of executing a pickup of a rider agent, but only when there are riders to be picked up.

The key to reproducing this pattern is that it must be part of a sudden and/or variable change scenario for arrival rates. We hypothesize that a stationary run will not emerge this phenomenon because, with unchanging arrival rates, incremental improvements will not allow for a critical mass of spatial imbalance in the location of riders and drivers.

There is a connected phenomenon observed in supply chain management theory that can be associated with this system property—what is known as the bullwhip effect. The phenomenon is widely understood as that of being a powerful reaction at the far end of a long supply chain which is often created from a small change in the point of origin of the chain. If the change is more sudden the effect is more compounded. In this case, the effect can be seen in the time-delayed spatial response of one agent group to another, not in a supply chain.

This behavior can be described as emergent. The pattern of maximizing utility through the prior, strategic positioning of driver agents is somewhat unexpected since all entry and exit of agents and their locations on the geography of the model are random. One might surmise (incorrectly, as we have shown) that if the random placement of rider agents and random placement of driver agents forms the core of the topological interactions of agents in this model, that through intuition alone there would be no clear gain in Voronoi movement behavior over random movement behavior. But as we have shown that while this is true for constant arrival rates, there is a difference in variable run conditions.

This emergent behavior does not seem to occur in the stationary runs of the model because at a constantly random and stable rate of entry for both agent types there is never a sudden vacuum to be capitalized upon.

This result provides researchers in this area with several important lessons: If one seeks to test movement behaviors of a ridesharing or an autonomous vehicle system, it should be tested under highly variable conditions in order to observe true emergent behavior. Stationary testing of spatial models would seem to be misleading and ineffective in this regard. Moreover, I propose that the testing of agent-agent interaction on any topological space for which there exists an entry or arrival dynamic would be subservient to the conclusions presented herein; though additional testing remains to conclude so irrefutably.

It's also paramount to realize that typical system dynamics modeling (differential equations) would most likely fail in producing the phenomenon as we have observed it, assuming a closed form equation for the mechanics of the model can be found, to begin with. Thus, we consider the experiment an added piece of important evidence for the utilization of agent-based techniques in autonomous vehicle and ridesharing service modeling and simulation (theoretically the difference between autonomous vehicle and ridesharing service modeling is negligible from the modeler's perspective.)

# 5. Summary & Future Work

In this paper, we described the development of an agent-based model for ridesharing services in the Washington, D.C. area. The model simulates riders and drivers through simple interactions on an accurate data-driven geospatial configuration. This model forms the basis for a number of experiments and model extensions that could yield greater insights into the ridesharing economy as it develops, expands, and evolves.

Our conclusions showed the importance of running experiments utilizing agent-based modeling runs not only in the form of stationary runs but in the form of variable scenario runs designed to create unpredictable effects that can—and in our case did—yield greater insights which otherwise would not have been observed.

Specifically, we found the emergence of a pattern where prior positioning of driver agents had a significant effect on pickup rates, and thus on the financial gain (utility) of drivers. We also found that this pattern emerged from a simple spreading-out behavior, which we called Voronoi movement and that this movement pattern outperformed random movement patterns even with randomly distributed arrival rates for both agent types. However, this emergent phenomenon was not observable unless a variable scenario was utilized in the experimentation process.

Consequently, we showed that driver to driver agent interactions, which form a symmetrical information environment can provide increased utility for drivers, and consequently for the ridesharing service and riders as well in some cases.

Current operating procedures of the leading ridesharing services do not allow drivers to gain access to location information of other drivers, and thus do not allow for movement behaviors that are dependent on that additional information. The symmetrizing of information can yield greater utility for all sides of this equation, including service, rider, and driver. Perhaps ridesharing services believe that giving less information to drivers would allow them more centralized control which they can use to better optimize the ridesharing experience, but evidence that this is true is not without question if we consider the natural fluctuations in drivers' and customers' supply and demand. The question posed by us here is whether more information may allow drivers to create adaptive and cooperative strategies to maximize their financial gain and by consequence, all other parties. We showed that this may be the case for one simple behavior and we demonstrated the conditions under which future spatial behavioral testing should be implemented if we are to be confident in the outputs of our simulations.

There are many pathways that this model can take—going forward. Primarily, the most interesting extension would be to add more spatially complex behaviors in agent-to-agent interactions and then to observe the results. It is not trivial that we test expected utility for an information symmetry scenario with only one movement-type behavior. More spatial movement patterns grounded in expected behaviors should be tested to quantify the difference in a service that allows more drivers to have more information and one that does not. There are improvements to be made in the spatial configuration of the model itself as well. For example, the inclusion of spatial demographics to enrich probability distribution calculations, adding road direction and traffic patterns, as well as utilizing more detailed geographic datasets would all make significant improvements to the model's efficacy and predictive power.

# References

1. Anderson, D. N. (2014). Not just a taxi: For-profit ridesharing, driver strategies, and VMT. *Transportation*, *41*(5), 1099–1117. doi:10.1007/s11116-014-9531-8
2. Bash, E. (2015). Dynamic Ridesharing: An Exploration of the Potential for Reduction in Vehicle Miles Traveled. Ph.D. Proposal.
3. Cho, S., Yasar, A. U. H., Knapen, L., Bellemans, T., Janssens, D., & Wets, G. (2012). A Conceptual Design of an Agent-based Interaction Model for the Carpooling Application. *Procedia Computer Science*, *10*, 801–807. doi:10.1016/j.procs.2012.06.013
4. Cho, S., Yasar, A.-U.-H., Knapen, L., Patil, B., Bellemans, T., Janssens, D., & Wets, G. (2013). Social networks in agent-based models for carpooling. In *Transportation Research Board 92nd Annual Meeting* (p. no. 13–2055).
5. Furuhata, M., Dessouky, M., F., Brunet, M. E., Wang, X., & Koenig, S. (2013). Ridesharing: The state-of-the-art and future directions. *Transportation Research Part B: Methodological*, *57*, 28–46. doi:10.1016/j.trb.2013.08.012
6. Hussain, I., Knapen, L., Galland, S., Yasar, A.-U.-H., Bellemans, T., Janssens, D., & Wets, G. (2015). Agent-based Simulation Model for Long-term Carpooling: Effect of Activity Planning Constraints. *Procedia Computer Science*, *52*(Ant), 412–419. doi:10.1016/j.procs.2015.05.006
7. Knapen, L., Keren, D., Yasar, A. U. H., Cho, S., Bellemans, T., Janssens, D., & Wets, G. (2012). Analysis of the co-routing problem in agent-based carpooling simulation. *Procedia Computer Science*, *10*(270833), 821–826. doi:10.1016/j.procs.2012.06.106
8. Knapen, L., Keren, D., Yasar, A.-U.-H., Cho, S., Bellemans, T., Janssens, D., & Wets, G. (2013). Estimating Scalability Issues While Finding an Optimal Assignment for Carpooling. *Procedia Computer Science*, *19*(Ant), 372–379. doi:10.1016/j.procs.2013.06.051
9. Martin, C. J. (2016). The sharing economy: A pathway to sustainability or a nightmarish form of neoliberal capitalism? *Ecological Economics*, *121*, 149–159. doi:10.1016/j.ecolecon.2015.11.027
10. Nielsen, J. R., Hovmøller, H., Blyth, P. L., & Sovacool, B. K. (2015). Of "white crows" and "cash savers:" A qualitative study of travel behavior and perceptions of ridesharing in Denmark. *Transportation Research Part A: Policy and Practice*, *78*, 113–123. doi:10.1016/j.tra.2015.04.033
11. Nourinejad, M., & Roorda, M. J. (2015). Agent-based model for dynamic ridesharing. *Transportation Research Part C: Emerging Technologies*, *1*, 117–132. doi:10.1016/j.trc.2015.07.016
12. Pelzer, D., Xiao, J., Zehe, D., Lees, M. H., Knoll, A. C., & Aydt, H. (2015). A Partition-Based Match Making Algorithm for Dynamic Ridesharing. *IEEE Transactions on Intelligent Transportation Systems*, *16*(5), 2587–2598. doi:10.1109/TITS.2015.2413453
13. Rayle, L., Shaheen, S., Chan, N., Dai, D., & Cervero, R. (2014). App-Based, On-Demand Ride Services: Comparing Taxi and Ridesourcing Trips and User Characteristics in San Francisco. *University of California Transportation Center*, *94720*(August), 1–20. doi:10.1007/s13398-014-0173-7.2
14. Rayle, L., Dai, D., Chan, N., Cervero, R., & Shaheen, S. (2016). Just a better taxi? A survey-based comparison of taxis, transit, and ridesourcing services in San Francisco. *Transport Policy*, *45*, 168–178. doi:10.1016/j.tranpol.2015.10.004
16. Shaheen, S. A. (2016). Shared mobility innovations and the sharing economy. *Transport Policy*, 1–2. doi:10.1016/j.tranpol.2016.01.008
17. Shaheen, S. A., Chan, N. D., & Gaynor, T. (2016). Casual carpooling in the San Francisco Bay Area: Understanding user characteristics, behaviors, and motivations. *Transport Policy*, 1–9. doi:10.1016/j.tranpol.2016.01.003

18. Tran, Q. T.; Tainar, D.; Safar, M. (2009). *Transactions on Large-Scale Data- and Knowledge-Centered Systems*. p. 357. ISBN 9783642037214.
19. Voronoi, G. (1908). Nouvelles applications des paramètres continus à la théorie des formes quadratiques. Premier mémoire. Sur quelques propriétés des formes quadratiques positives parfaites. *Journal Für Die Reine Und Angewandte Mathematik (Crelle's Journal)*, *1908*(133). doi:10.1515/crll.1908.133.97
20. Wilensky, U. (1999). NetLogo. http://ccl.northwestern.edu/netlogo/. Center for Connected Learning and Computer-Based Modeling, Northwestern University, Evanston, IL
21. Zargayouna, M., Zeddini, B., Scemama, G., & Othman, A. (2013). Agent-based simulator for travelers multimodal mobility. *Frontiers in Artificial Intelligence and Applications*, *252*, 81–90. doi:10.3233/978-1-61499-254-7-81
22. Zhang, W., Guhathakurta, S., Fang, J., & Zhang, G. (2015). Exploring the impact of shared autonomous vehicles on urban parking demand: An agent-based simulation approach. *Sustainable Cities and Society*, *19*, 34–45. doi:10.1016/j.scs.2015.07.006
23. Zolnik, E. J. (2015). The effect of gasoline prices on ridesharing. *Journal of Transport Geography*, *47*, 47–58. doi:10.1016/j.jtrangeo.2015.07.009